\documentclass{aa}
\usepackage{graphicx}
\usepackage{natbib}
\usepackage[varg]{txfonts}
\begin{document}
   \title{The chromosphere above sunspots at millimeter wavelengths}

   \author{M. Loukitcheva
          \inst{1,2}
          \and
          S. K. Solanki \inst{1,3}
          \and
          S. M. White \inst{4}
          }

   \institute{
      Max-Planck-Institut for Sonnensystemforschung, D-37191 Katlenburg-Lindau, Germany\\
    \email{lukicheva@mps.mpg.de}
    \and
    Astronomical Institute, St.Petersburg University, Universitetskii pr. 28, \\ 198504 St.Petersburg, Russia
    \and
    School of Space Research, Kyung Hee University, Yongin, Gyeonggi 446-701, Korea
                 \and
    Space Vehicles Directorate, Air Force Research Laboratory,
Kirtland AFB, NM, United States
             }

   \date{Received ; accepted }


  \abstract
   {}
   {The aim of this paper is to demonstrate that millimeter wave data can be used to distinguish between
various atmospheric models of sunspots, whose temperature
structure in the upper photosphere and chromosphere has been
the source of some controversy.}
   {We use observations of the temperature contrast (relative to the quiet Sun) above a
sunspot umbra at $3.5$~mm obtained with the Berkeley-Illinois-Maryland
Array (BIMA), complemented by sub-mm observations from \citet{lindsey}
and 2 cm observations with the Very Large Array. These are compared with the umbral contrast calculated from various atmospheric models of sunspots.}
   {Current mm and sub-mm observational data suggest that the brightness observed
at these wavelengths is low compared to the most widely
used sunspot models. These data impose strong constraints on the temperature and density stratifications
of the sunspot umbral atmosphere, in particular on the location and depth of
the temperature minimum and the location of the transition region.}
   {A successful model
that is in agreement with millimeter umbral brightness should have an
extended and deep temperature minimum (below 3000~K). Better spatial resolution as well as better wavelength coverage
are needed for a more complete determination of the chromospheric temperature stratification above sunspot umbrae.
}


 \maketitle
%

\section{Introduction}

There have been numerous attempts to build a comprehensive
semiempirical model of the atmosphere above a sunspot umbra. Deriving such
models is a complicated process that tries to balance observations of a
range of optical and UV lines \citep[mostly formed in non-LTE conditions, see,
e.g.,][for a review see Solanki 2003]{Avrett1981, Maltby1986, Obridko}
and, when available, radio measurements of brightness
spectra, together with ionization equilibrium and radiative transfer
calculations that include heat transfer down from the corona as well as
other factors \citep[e.g.,][]{Fontenla1993}.  The radio data are
particularly valuable for solar diagnostics because the measurements are in
the Rayleigh--Jeans limit, meaning that measured brightness temperatures
actually represent thermal electron temperatures in the optically thick
atmosphere. The temperature, at which a given frequency is optically thick,
is sensitive to density and temperature, and hence the radio data provide
important constraints for modeling \citep[e.g.,][]{loukitcheva}.

However, while there are numerous radio temperature measurements of the
quiet--Sun atmosphere, there are very few for sunspots, particularly at
millimeter wavelengths, since good spatial resolution is required to
isolate the brightness temperature of the sunspot from its surroundings, and single--dish
measurements generally do not have enough resolution. The
highest--resolution single--dish data, at $10-20$\arcsec, are from JCMT
\citep{lindsey} and CSO \citep{bastian}, but at sub-mm
wavelengths, which sample deeper in the atmosphere, closer to photospheric
heights. The sunspot umbra
appears darker than the quiet Sun at these wavelengths and umbral contrast
decreases towards longer wavelengths. It is more interesting,
however, to use millimeter wavelength radio data, which can constrain the
temperature in the chromosphere and address the controversy regarding the umbral
temperature in these layers \citep[see, e.g.,][and Sect.~\ref{section3}]{Maltby1986, Severino, Fontenla2009}.  At millimeter
wavelengths the instrument best suited for high-resolution observations of
sunspots until recently was the 10-element Berkeley-Illinois-Maryland Array
(BIMA) operating at $3.5$~mm \citep{welch}. The BIMA antennas are
now part of the Combined Array for Millimeter Astronomy \citep[CARMA;][]{Bock2006}.

 \begin{figure*}[!htb]
   \centering
   \includegraphics[width=0.32\textwidth, angle=90]{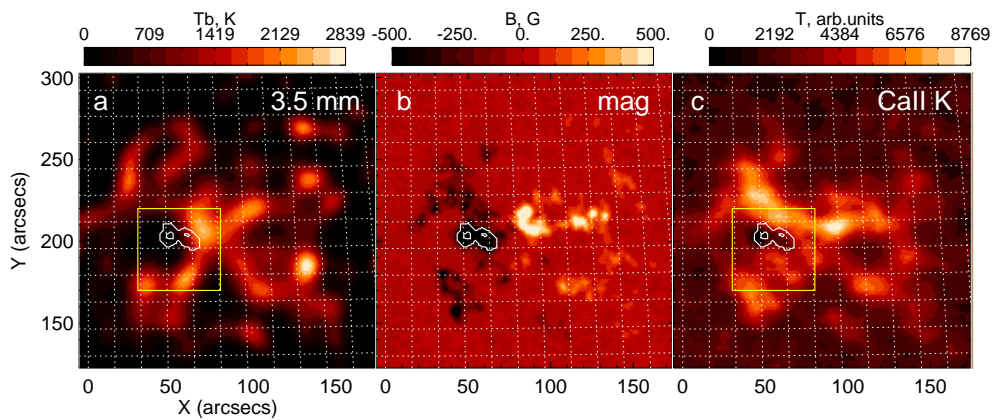}
      \caption{Active region NOAA 10448 observed on August 31,
        2003. \textit{(a)} BIMA image at 3.5 mm, \textit{(b)} MDI
        photospheric magnetogram and \textit{(c)} BBSO Ca~{\sc ii}~K image. White
        contours mark sunspot umbrae and penumbrae. The box contains the
        region shown in Fig.~\ref{fig3} as a blowup. The zero brightness in the BIMA image corresponds to the
weakest flux.}
         \label{fig1}
   \end{figure*}

\begin{figure}[!htb]
   \centering
   \includegraphics[width=0.25\textwidth, angle=90]{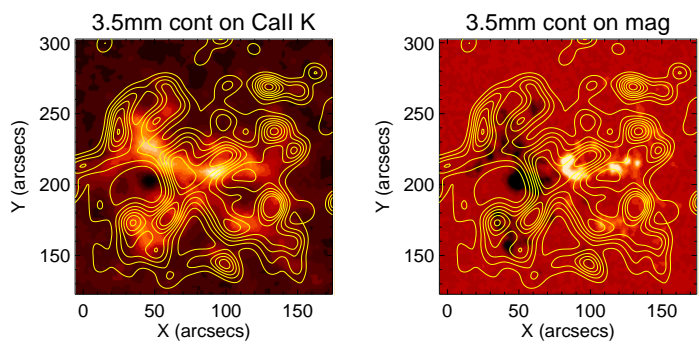}
      \caption{BBSO Ca~{\sc ii}~K image and MDI
        photospheric magnetogram with the overlaid contours of 3.5 mm emission plotted as
        iso-intensity lines corresponding to 10, 20,
        30, 40, 50, 60, 70 \% of maximum brightness. }
         \label{fig22}
   \end{figure}

\begin{figure}[!htb]
   \centering
   \includegraphics[width=0.25\textwidth, angle=90]{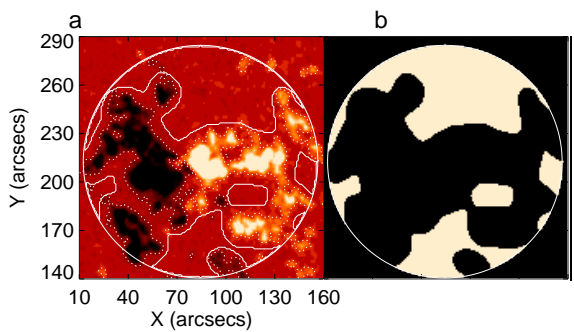}
      \caption{\textit{(a)} MDI magnetogram in the range
        (-200,200)~Gauss. The dotted and solid lines mark the 40~G and 26~G
        contours, corresponding to the quiet-Sun level at the original
        resolution of the magnetogram and at the resolution of the mm data,
        respectively. The white
        circle corresponds to the BIMA primary beam of 3\arcmin\ by
        3\arcmin\ containing flux.\textit{(b)} The resulting quiet-Sun mask restricted to BIMA's primary beam.
        Quiet-Sun areas are white. }
         \label{fig2}
   \end{figure}

In this work we use BIMA observations of the temperature contrast (relative to the quiet Sun) above a
sunspot umbra at $3.5$~mm. At this wavelength models predict rather different umbral brightnesses, so that such measurements can be used to
distinguish between the various atmospheric models of sunspots. In
Section~\ref{section2} we report the observations of the active region at
$3.5$~mm, evaluate umbral brightness at this wavelength relative to the
quiet Sun, and investigate the influence of spatial resolution on the
appearance of the sunspot at mm wavelengths. In Section~\ref{section3} we discuss
the differences between the temperature and density stratifications in the
existing sunspot models as well as between mm brightness spectra calculated
from these models. We demonstrate that observations of sunspots at mm
wavelengths impose strong constraints on temperature and density
stratifications of the sunspot atmosphere, particularly on the location and
depth of the temperature minimum and the location of the transition region
in the umbral models. In Section~\ref{section4} we complement the investigation of sub-mm and mm umbral brightnesses
with two examples of the observations of sunspot umbra at short
cm wavelengths.

\section{Observational data and their analysis}\label{section2}

On August 31, 2003 BIMA observed a small active region NOAA 10448 north of
the solar equator (coordinates N20W07 on 31-Aug-2003 23:30 UT). The BIMA images
were deconvolved using the maximum entropy method (MEM) and restored with a
Gaussian beam of 12\arcsec\ \citep{white}. Contemporaneous images of
NOAA 10448 at 3 wavelengths are shown in Fig.~\ref{fig1} including the BIMA
image at 3.5 mm wavelength, a photospheric magnetogram from the Michelson
Doppler Imager (MDI) on the SOHO satellite \citep{Scherrer}, and a Ca~{\sc ii}~K image obtained
at the Big Bear Solar Observatory. Prior
to the analysis solar differential rotation was compensated, so that all image locations correspond to the AR's location at a single time. Then the images were spatially
coaligned. In Fig.~\ref{fig22} we show the results of the coalignment depicting the mm emission contours overlaid on the Ca~{\sc ii}~K image and the magnetogram.
The pixel size for all images was set to 1\arcsec. Images are
241 by 241 pixels which corresponds to a 241\arcsec\ by 241\arcsec\ FOV.

 \begin{figure*}[!htb]
   \centering
   \includegraphics[width=0.32\textwidth,angle=90]{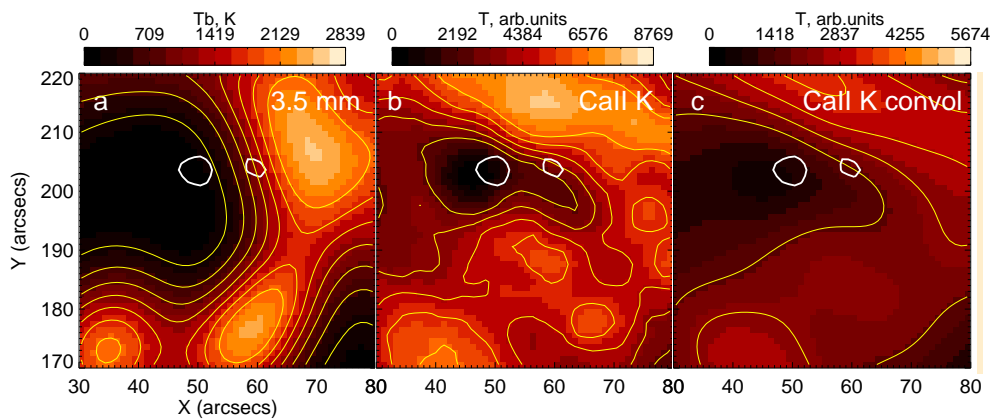}
      \caption{A blowup of the sunspot region marked in
        Fig.~\ref{fig1}. \textit{(a)} BIMA image at 3.5 mm, \textit{(b)}
        BBSO Ca~{\sc ii}~K image at the original resolution and \textit{(c)} the
        Ca~{\sc ii}~K image convolved to the resolution of the mm image
        (12~\arcsec). Thick white contours mark the locations of two
        umbrae. The other are iso-intensity lines corresponding to 10, 20,
        30, 40, 50, 60, 70 \% of maximum brightness for the mm image and
        30, 40, 50, 60, 70 \% of maximum brightness for the original and
        convolved calcium images. The zero brightness in the BIMA image corresponds to the
         weakest flux.}
         \label{fig3}
   \end{figure*}

For analysis and comparison with the umbral models we chose the largest
sunspot in the field, which is located at the trailing (eastern) edge of
the active region. Its photospheric magnetic field does not exceed 2000~G
according to recalibrated MDI full disk magnetograms\footnote{However,
it is likely that the recalibrated magnetograms still significantly
underestimate the total flux in the sunspot: see \citet{Ulrich}.}.
This sunspot had a well--defined umbra and
penumbra, with the penumbra linking to a smaller sunspot with a
weaker umbra to the west of the larger spot (see corresponding umbral and
penumbral contours
overlaid in Fig.~\ref{fig1}). Hereafter we refer to the umbra under
consideration, associated with the largest sunspot, as ``the big'' umbra,
and for comparison we usually provide the observational quantities obtained for
``the small'' umbra of the western sunspot as well. The umbral and penumbral
boundaries were derived from the MDI continuum image as, respectively, $0.65$ and $0.90$
of the surrounding photospheric white--light intensity following \citet{Mathew}.

Due to a positivity constraint applied in the maximum entropy deconvolution
method the zero level in the BIMA maps corresponds to the
weakest flux in the image. Therefore, to determine umbral brightness
decrement relative to the QS we first need to establish the quiet Sun level in
the BIMA images. To that end we used the MDI/SOHO photospheric magnetograms
and applied a threshold of 40 G in the absolute value of the magnetogram
signal to determine quiet-Sun regions.  After that we degraded the
resolution of the (absolute--valued) magnetogram to the BIMA resolution
(12\arcsec) and found the contour carrying the same amount of magnetic flux
as the 40~G contour in the magnetogram at the original MDI resolution
(2.6\arcsec). The
resulting quiet-Sun brightness was determined as the average of brightness
outside this iso-Gauss contour, which turns out to lie at 26~G at the
12\arcsec\ resolution of the mm data.
The results of the procedure are demonstrated in Fig.~\ref{fig2}a, where we
show the magnetogram with the overlaid 40~G and 26~G contours, represented by
dotted and solid lines, respectively. The latter corresponds to the
threshold defining the resulting quiet-Sun mask, depicted in
Fig.~\ref{fig2}b. Note that we consider the flux only within the BIMA
field of view of 3\arcmin\ by 3\arcmin\ (white circle in Fig.~\ref{fig2}).

We acknowledge that there is considerable uncertainty in
choosing the pixels that correspond to the ``quiet--Sun atmosphere'', since
there is always some arbitrariness in the criterion chosen to identify
such pixels.  The criterion applied here is
consistent with other investigations  \citep[e.g.,][]{Wang} and
leads to a reasonable proportion of pixels in quiet Sun regions. Using
even stronger thresholds (based on the mm brightness histogram) we would get
lower values for the level of quiet--Sun emission in the
millimeter images, but they are always higher than the derived umbral
brightness.

   \begin{figure}
   \centering
   \includegraphics[width=0.45\textwidth]{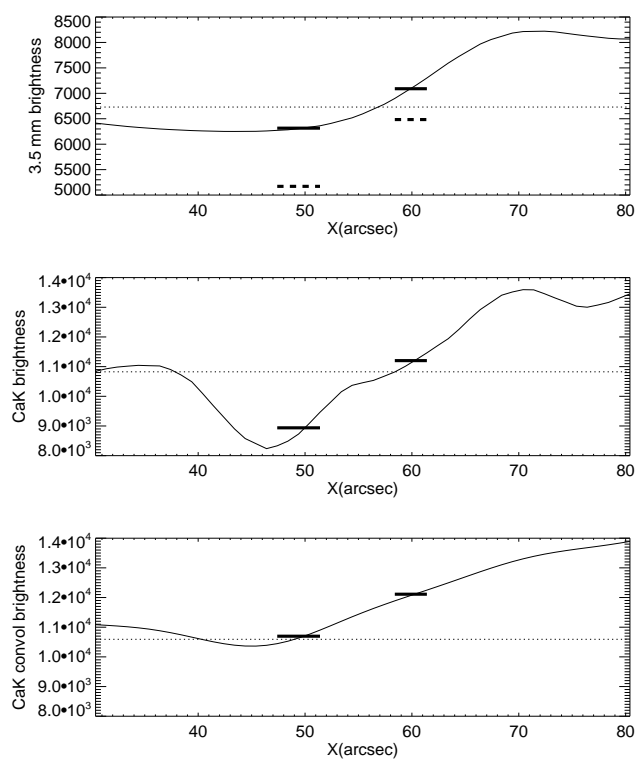}
      \caption{The horizontal cuts through the umbrae at
        $y=203.6$\arcsec\ north of apparent solar disk center for BIMA image
        at 3.5 mm \textit{(a)}, for BBSO Ca~{\sc ii}~K image at the original
        resolution \textit{(b)} and at 12~\arcsec\ resolution of the mm image \textit{(c)}. Thick solid lines mark the
        location and size of the two umbrae. Thick dashed lines correspond
        to the potential reduction of the mm umbral brightness when going
        to 4\arcsec\ resolution. The QS level is indicated by the thin dotted
        lines.}
         \label{fig4_1}
   \end{figure}

A high level of correspondence between the mm brightness contours and the magnetic field structures, as well as between the mm brightness contours and calcium intensity, found in Fig.~\ref{fig22}, indicates that the mm brightness decrement at the location of the umbrae can provide an estimate of umbral brightness at 3.5 mm even if there is no obvious umbral feature found in the mm image. We calculate the umbral millimeter brightness as the average brightness of
pixels lying within the umbral boundaries. Special attention was paid to
the fact that in the mm image the location of the weakest emission (the
zero level in the maximum entropy maps) does not match the exact position
of the umbra, while the darkest feature in the calcium image lies within the
umbral boundary. \citet{lindsey} also reported that the location
of dark features corresponding to sunspot umbrae in the sub-mm images often
appeared to be noticeably offset
from their locations in the Ca~K images. To study the
influence of image resolution on the appearance of the umbral depression and
its position relative to the photospheric umbral counterpart we convolved
the Ca~{\sc ii}~K image with the 12\arcsec\ beam corresponding to the BIMA
resolution and plotted a blowup (marked with a box in Fig.~\ref{fig1}) of
the convolved image together with the calcium image at the original
resolution and the 3.5~mm image in Fig.~\ref{fig3}. In the calcium image,
smoothed to the resolution of the mm image (Fig.~\ref{fig3}c), the
brightness minimum is no longer within the umbral boundaries and the
depression in brightness becomes less pronounced and more diffuse
due to an
admixture of signal from neighboring bright points. Therefore smearing of the
brightness distribution reduces umbral contrast relative to the quiet-Sun
background. This result is demonstrated nicely in Fig.~\ref{fig4_1}, where
we show the east--west cuts through the two spot umbrae at
$y=203.6$\arcsec\ north of apparent disk center for the mm image and
calcium images
at original and degraded (12\arcsec) resolution respectively. In
Fig.~\ref{fig4_1} we adopted for convenience $T_{b}=6730$K as the absolute
quiet Sun level at 3.5~mm as an
average of the several single--dish quiet-Sun brightness temperature
measurements \citep{Nagnibeda,loukitcheva2005}.

For the big umbra (which is used further for comparison with umbral models below)
the average calcium umbral contrast changes from $0.83$ of the quiet-Sun
brightness in the original image to $1.04$ in the smoothed image (see
Fig.~\ref{fig4_1}b and Fig.~\ref{fig4_1}c), i.e. the big umbra in the degraded Ca images is slightly brighter than the QS, mainly due to admixture of bright emission from nearby plage. For the small umbra located in
the region of the enhanced calcium flux the values are $1.01$ and $1.14$
respectively.  Degrading the image resolution also decreases the
contrast between the big and the small umbrae. The ratio of umbral
brightnesses changes from $0.80$ in the original image to $0.88$ in the
smeared image. The latter value is close to the ratio derived from the mm
image, which is $0.90$ (Fig.~\ref{fig4_1}a). Therefore the results shown
in Figs.~\ref{fig3} and \ref{fig4_1} confirm that the relatively low resolution of
the mm images leads to an overestimate of the umbral brightness and is a
plausible reason for the displacement of the location of the minimum
in the mm image relative to its photospheric umbral counterpart.

The distribution of brightness in the mm and degraded calcium images is
sufficiently different that we cannot use the comparison between original
and degraded calcium images to correct the BIMA umbral brightness, but we
can use it to estimate the uncertainty in the mm umbral contrast relative
to the QS. Thus, in the upper panel of Fig.~\ref{fig4_1} we plot (thick
dashed lines) the potential reduction of the umbral brightness at 3.5~mm
when going to 4\arcsec\ resolution if we assume
the same contrast between the umbrae and the QS for the mm and calcium
images. The same effect is shown also in Fig.~\ref{fig5} with downward
arrows for the mm brightness of the big and small umbrae depicted with
filled and open circles, respectively. We estimate this uncertainty to be 17\% and 9\%
of the QS brightness at 12\arcsec\ resolution for the big and small umbrae, respectively.
Note that this correction can be considered to be an upper limit, so that the true contrast is expected to lie between thick solid and dashed bars in Fig.~\ref{fig4_1}.

From the above analysis of the BIMA observations,
we estimate the ``3.5 mm umbra'' to
be approximately $400$~K cooler than the quiet sun
(roughly 6\% of the QS brightness temperature at this wavelength) at the
resolution of 12\arcsec\ with an uncertainty of order
$100$~K.  This is the value we will use primarily for comparison with
the atmospheric models, but in addition we will also reference the
resolution--corrected (based on Ca {\sc ii}) value of $1500$~K as the upper
limit to the umbral decrement at 3.5 mm. For comparison with the model calculations we also use the sub-mm
umbral brightness temperatures measured at $0.35$~mm, $0.85$~mm, and $1.2$~mm from \citet{lindsey}.
The observed sub-mm and mm umbral brightness and corresponding measurement
error estimates are plotted in Figs.~\ref{fig5} and \ref{fig10} together with the output of the
sunspot models discussed in the following section.


\section{Model simulations}\label{section3}

In order to distinguish between the sunspot models, we have calculated the
expected sub/mm brightness temperatures at $24$ selected wavelengths in the
range $0.1-20$~mm for a set of classical sunspot models as well as their
more recent updates. The models we analyzed are the sunspot model of \citet{Avrett1981}, model $M$ of \citet{Maltby1986}, the sunspot model of \citet{Severino}, models $A$ (dark umbra) and $B$ (bright umbra) of \citet{socas}, and model $S$ of \citet{Fontenla2009}. 
The first two models are
classical semiempirical sunspot models covering the umbral photosphere,
chromosphere and transition region. The model of \citet{Severino} is
an update of the model of \citet{Caccin} which is in its turn a
modified version of one of the \citet{Maltby1986} models with a steeper
temperature gradient in the photosphere. The model of \citet{socas}
was derived from non-LTE inversions of high-resolution spectropolarimetric
observations of four Ca~{\sc ii} and Fe~{\sc i} lines. The most recent model of \citet{Fontenla2009} 
is based on the EUV spectrum from the
SOHO/SUMER atlas \citep{Curdt}. We complemented this set of the one-component umbral models
with the two-component model of \citet{Obridko}. Their umbral
atmosphere consists of the main (dark) component and a secondary (bright)
component with a filling factor in the range $0.05-0.1$ depending on the
spot size and solar cycle phase.

\begin{figure}[!htb]
   \centering
  \includegraphics[width=0.45\textwidth]{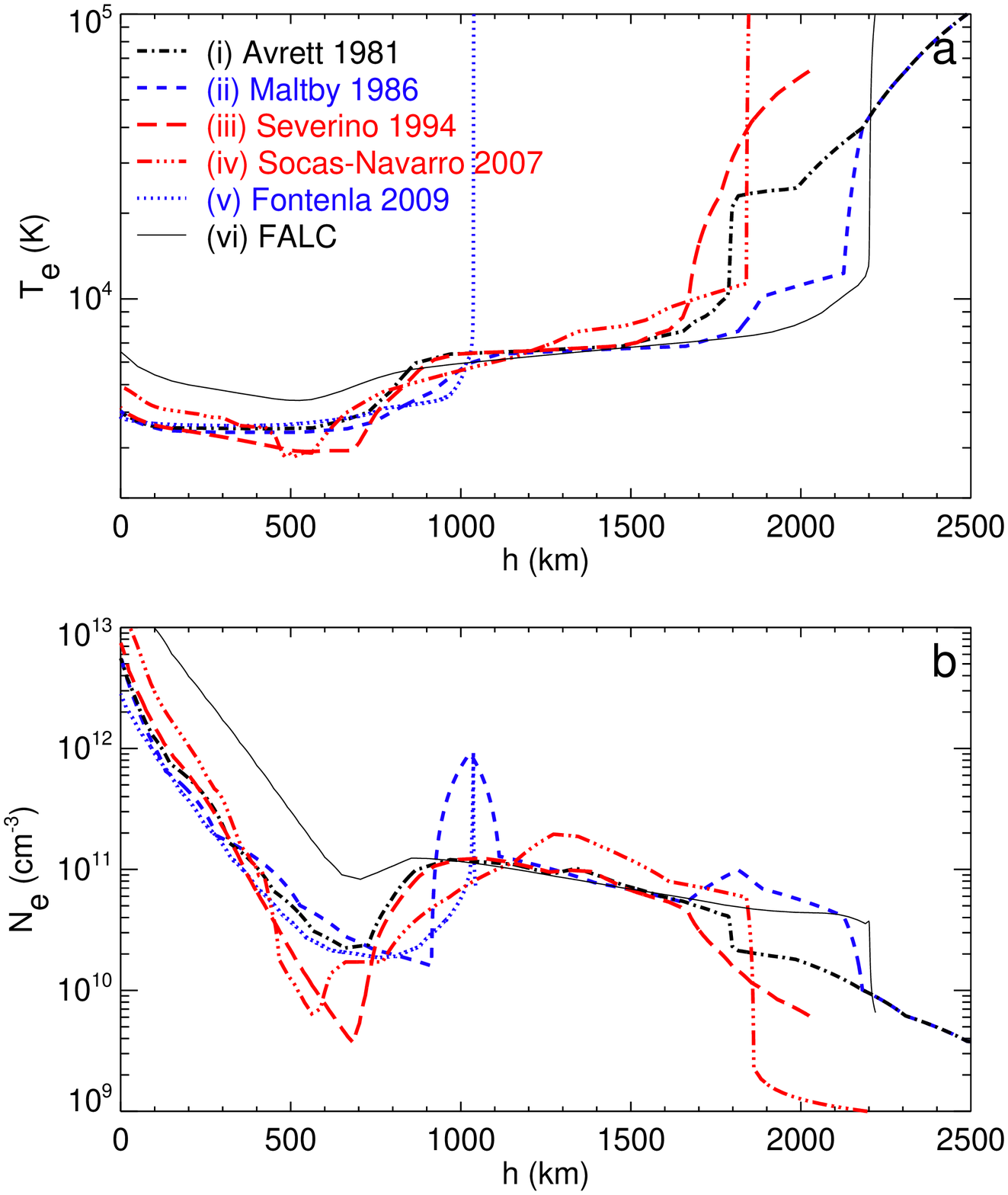}
      \caption{\textit{(a)} Electron temperature as a function of height in
        a number of standard models of the solar chromosphere above a
        sunspot umbra. (i) the model of \citet{Avrett1981}, (ii) of \citet{Maltby1986}, (iii) of \citet{Severino},
        (iv) of \citet{socas}, (v) of \citet{Fontenla2009}.
        The thin solid black line (vi) is the reference quiet--Sun atmosphere of
        \citet{Fontenla1993} known as $FAL-C$. \textit{(b)}
        The electron number density as a function of height for the same
        models as shown in panel (a).}
         \label{fig4}
   \end{figure}

The height dependence of the electron temperature and electron number density
of each of these models are plotted in Fig.~\ref{fig4}, together with the
reference quiet--Sun atmosphere. As a reference model for this plot we
chose the model that represents the average quiet Sun (model $C$ of
\citet{Fontenla1993}, commonly referred to as $FAL-C$). We preferred it to
the more recent quiet-Sun models of \citet{Fontenla2007} and of \citet{Avrett2008}.
Although the quiet-Sun model of \citet{Fontenla2007} is
constructed at moderate resolution it represents not the average quiet Sun
but the internetwork. Furthermore, the model parameters in the upper
chromosphere and transition region in this model are described as tentative
(see Section 3 of \citet{Fontenla2007} and consequently the model cannot account accurately for the
formation of the mm emission at the spatial resolution achieved by BIMA. The model of \citet{Avrett2008}, an
updated average quiet-Sun model, produces millimeter brightness temperatures
substantially lower than any of the observed values \citep[see Fig.3 of][]{Avrett2008}.

We note that each of the the umbral models analyzed in this paper
compared their results with corresponding quiet-Sun atmosphere models,
but that different umbral models used different quiet--Sun models as
their reference. For instance, the average QS
model by \citet{Vernazza}, $VAL-C$, was used as a reference point by
\citet{Maltby1986} and \citet{Obridko}. The model by \citet{Gingerich}, HSRA model, was chosen to represent the QS in the umbral models of \citet{socas}.

Calculations of the expected brightness temperature as a function of
wavelength in the mm/\-submm range were carried out under the assumption that
bremsstrahlung opacity
is responsible for the mm continuum radiation. We distinguish two sources
of opacity: due to encounters between free electrons and protons, and between
free electrons and neutral hydrogen. They are
usually referred to as $H^0$ opacity and $H-$ opacity, respectively. We
assume that gyroresonance opacity is negligible at these high
frequencies. The expression for bremsstrahlung opacity in magnetic media
was taken from \citet{zheleznyakov}. The magnetic field was modelled by a
vertical dipole buried under the photosphere, following the approach of
\citet{zlotnik}. For the calculations we adopted the value of 2000~G for
photospheric umbral magnetic field strength and a penumbral size (outer
diameter) of 12\arcsec, which corresponds to a dipole depth of 10000~km.

From Fig.~\ref{fig4} it is seen that the sunspot models differ
significantly from the $FAL-C$ quiet-Sun model and from each other in the
depth and extension of the temperature minimum region and also in the
location of the transition region. A number of classical models, such as
\citet{Avrett1981}, marked in Fig.~\ref{fig4} with (i), \citet{Maltby1986} (ii), and also their update by \citet{Severino} (iii), and the model of \citet{socas} (iv),
have the transition region at heights similar to
the $FAL-C$ model (about 2000~km and above). However, the more recent
model of \citet{Fontenla2009} (v) 
places the transition region at much lower heights (around 1000~km).  The
temperature minimum in the sunspot umbra models generally lies deeper than
in the $FAL-C$
model. Furthermore the temperature in the temperature minimum
region drops to below 3000~K in the models of \citet{Severino} (iii) and
of \citet{socas} (iv). However, the model of \citet{Severino} (iii)
possesses an extended temperature minimum, i.e. the temperature minimum reaches to much greater heights in the atmosphere than do those of the other models.

\begin{figure}[!htb]
   \centering
   \includegraphics[width=0.38\textwidth]{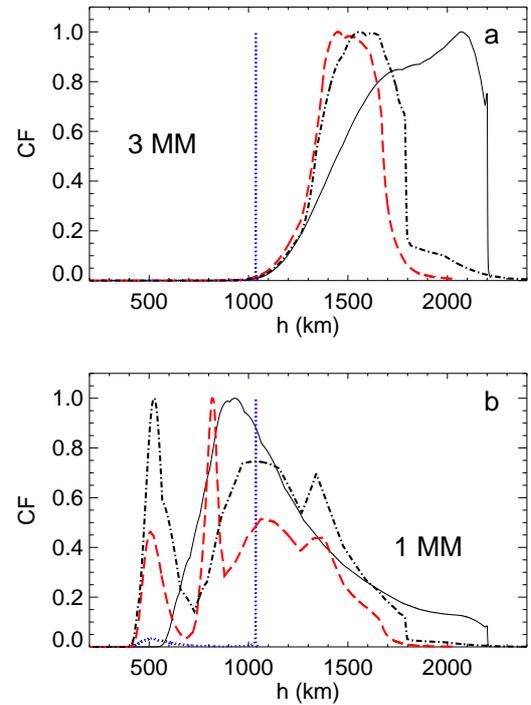}
      \caption{Brightness--temperature contribution functions (with maxima normalized to unity) at 3~mm
        \textit{(a)} and at 1~mm \textit{(b)} for a number of sunspot
        models including the model of \citet{Avrett1981} (dot-dashed curve), of
        \citet{Severino} (dashed curve), and of \citet{Fontenla2009}
        (dotted curve). 
        For comparison
        the contribution function for the model $FAL-C$ is shown with a thin
        solid black line.}
         \label{fig6}
   \end{figure}

To study the contribution of various atmospheric layers to the emergent
intensity of the radiation we calculated the brightness temperature
contribution functions for the sunspot umbral models introduced in
Fig.~\ref{fig4}. In Fig.~\ref{fig6} we show the contribution functions at
3~mm and 1~mm for the umbral models of \citet{Avrett1981}, \citet{Severino},
and \citet{Fontenla2009}. For comparison we
also plot the contribution function for model $FAL-C$ with a thin solid
line in Fig.~\ref{fig6}.

The models of \citet{Avrett1981} and \citet{Severino} both attribute
the major contribution to radiation at $3$~mm wavelength
to a similar range of heights, of
order 1200--1800~km, and moreover the contribution functions are of similar shape
(see corresponding dot-dashed and dashed curves in Fig.~\ref{fig6}a). The
reason for this is the similarity of the electron temperature and
number density distributions for these two models in this range of heights
(see (i) and (iii) in Fig.~\ref{fig4}). In the model of \citet{Fontenla2009} (dotted line in Fig.~\ref{fig6}a)
the effective emitting region at 3~mm is sharply defined due to the
fact that the coronal contribution is negligible and the absorption coefficient
varies rapidly with height at the base of the (low-lying)
transition region. 

Going to shorter wavelengths the formation heights move lower in the
chromosphere (see Fig. \ref{fig6}b for 1~mm). However, for the model of
\citet{Fontenla2009} the only difference between the contribution
functions at 3 and 1~mm is a small bump of additional contribution from
heights close to 500~km. The other models also exhibit a
substantial contribution from heights at around
500~km. At these heights all the models investigated provide favorable
conditions, namely low temperatures together with sufficiently high
neutral hydrogen and electron densities for $H-$ opacity to become
appreciable,
which leads to the observed peaks in the contribution functions. The
contribution function (dashed line in Fig.~\ref{fig6}b) for the model of
\citet{Severino} is characterized by an additional peak at 800~km
which is also due to effective collisions between electrons and neutral
hydrogen at this height.

Thus, analysis of the contribution functions for
different umbral models demonstrates that at 3~mm we should sample mostly
the middle and upper umbral chromosphere, but in the model of \citet{Fontenla2009}
the transition region also contributes significantly due to its location
low in the atmosphere. For radiation at 1~mm
and shorter wavelengths a substantial contribution comes from the heights of the
temperature minimum region.

   \begin{figure}[!htb]
   \centering
   \includegraphics[width=0.35\textwidth, angle=90]{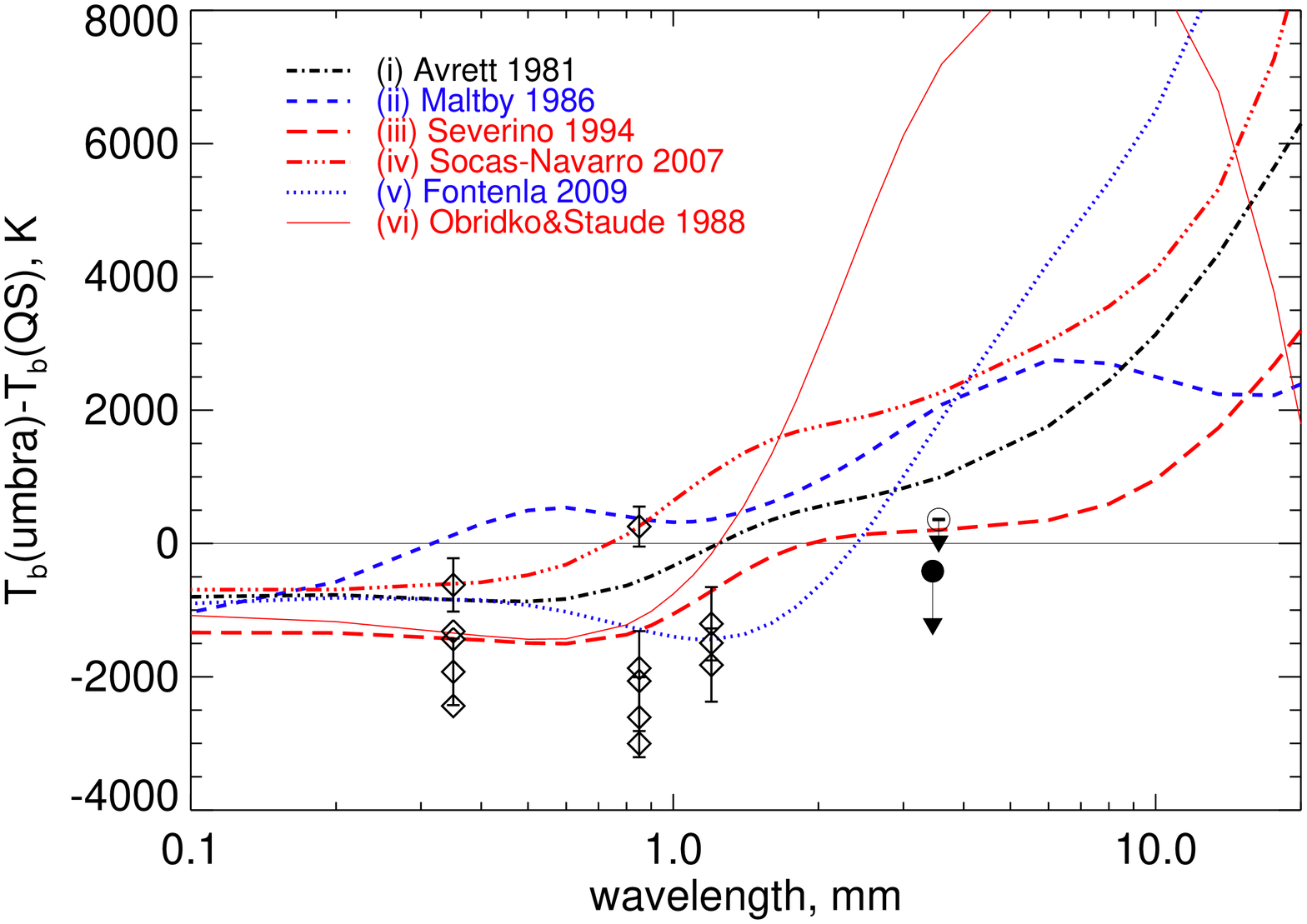}
      \caption{The difference between the umbral brightness (in temperature units) and the QS brightness, plotted as a function of
        wavelength for the model (i) of \citet{Avrett1981}, (ii) model $M$ of
        \citet{Maltby1986}, (iii) of \citet{Severino}, (iv) model
        $A$ of \citet{socas}, (v) model $S$ of \citet{Fontenla2009} and
        (vi) the two-component umbral
        model of \citet{Obridko} employing a filling
        factor of $0.05$ for the hot component. Filled and open circles mark the observational
        values obtained from BIMA maps at 3.5 mm for the big and small
        umbrae, respectively. Arrowheads mark the shift of umbral
        brightness when extrapolating to 4\arcsec\ resolution. Diamonds
        stand for the measurements from JCMT at 0.35, 0.85 and 1.2 mm made by \citet{lindsey}.}
         \label{fig5}
   \end{figure}

    \begin{figure}[!htb]
   \centering
    \includegraphics[width=0.35\textwidth, angle=90]{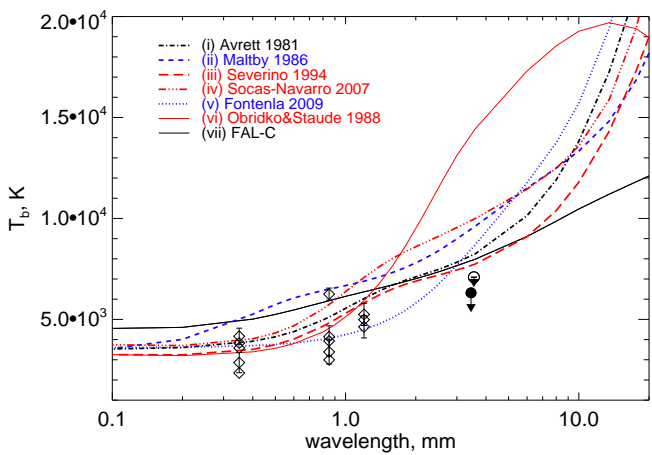}

      \caption{The absolute brightness temperature as a function of
        wavelength for the same umbral models as in Fig.~\ref{fig5}. Also plotted is
        (viii) $FAL-C$ model representing the Quiet Sun. }
         \label{fig10}
   \end{figure}

Figures~\ref{fig5} and \ref{fig10} show the predicted umbral brightness temperature spectra
from each model, together with the BIMA observational results and
measurements culled from \cite{lindsey} depicted by circles and diamonds,
respectively. In Fig.~\ref{fig5} we consider each umbral model relative to the QS model that it refers to and plot the corresponding mm brightness contrast $T_b^{umbra} - T_b^{QS}$. We chose to depict the difference between the umbral and QS brightness in order to minimize the influence of the reference QS value on the umbral brightness and to make the models directly comparable with the observations which provide only the difference of the umbral brightness relative to the QS. For completeness, in Fig.~\ref{fig10} we compare the absolute brightness spectra from the umbral models with that from a single standard quiet-Sun model which is represented by the model $FAL-C$ (plotted with black solid line in Fig.~\ref{fig10}).

From Figs.~\ref{fig5} and ~\ref{fig10} it is seen that the models by \citet{Severino} and \citet{Fontenla2009} reproduce the majority of the submm-mm observational data from \citet{lindsey}
relatively well. These models are characterized by deep and extended
temperature minima (see Fig.~\ref{fig4}) which dominates the
contribution functions for
umbral brightness at wavelengths up to 1~mm
(Fig.~\ref{fig6}b). However at 3.5~mm none of the models displays brightness in the umbra that is reduced relative to the individual
reference quiet--Sun models (Fig.~\ref{fig5}), in contrast to the observations which find
at least the large sunspot umbra to be dark at this wavelength. Rather, these models predict
that the umbra should be brighter than the surrounding quiet Sun at
3.5~mm wavelength.
The model of \citet{Severino} predicts the lowest 3.5~mm
umbral brightness and is therefore the closest to the observed value. In this
model a wide range of heights in the upper chromosphere (1300-1700~km)
contributes to the radiation at 3.5~mm wavelength (Fig.~\ref{fig6}a) and a
better fit to the observed values might need a revision of the temperature
and electron density stratifications at these heights.
The sub-mm and mm brightnesses predicted by the
classical umbral models of \citet{Avrett1981} and \citet{Maltby1986}, as well as the
dark umbral model of \citet{socas} and the two-component model of
\citet{Obridko}, appear to be too high to account for the current
submm-mm observations.

We expect the true
3.5~mm umbral brightness to be lower than the
reported value of 400~K below the QS level (filled circle in
Fig.~\ref{fig5}) due to the dilution of the brightness--temperature
reduction by the relatively low resolution of the data and also show
in Figs.~\ref{fig5} and \ref{fig10} the potential 3.5~mm brightness extrapolated (albeit crudely, based on Ca~{\sc ii}~K images) to a
resolution of 4\arcsec\ with a downward arrow.

    \begin{figure}[!htb]
   \centering
  \includegraphics[width=0.35\textwidth, angle=90]{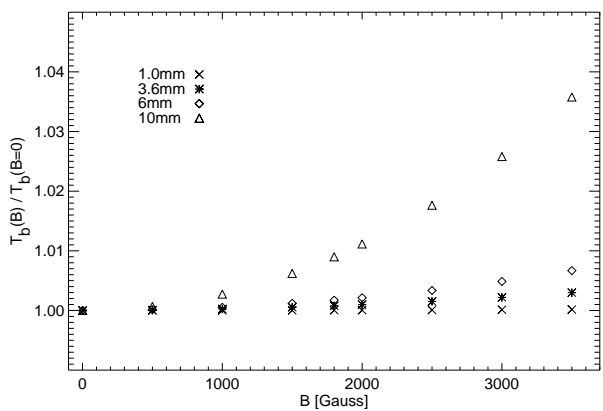}
      \caption{Normalized brightness temperature as a function of sunspot
        photospheric magnetic field strength for the model of \citet{Fontenla2009} for 1~mm (cross), 3.6~mm (asterisk), 6~mm (diamond) and
        10~mm (triangle).}
         \label{fig7}
   \end{figure}

\begin{figure*}[!htb]
 \centering
 \includegraphics[width=0.7\textwidth, angle=0]{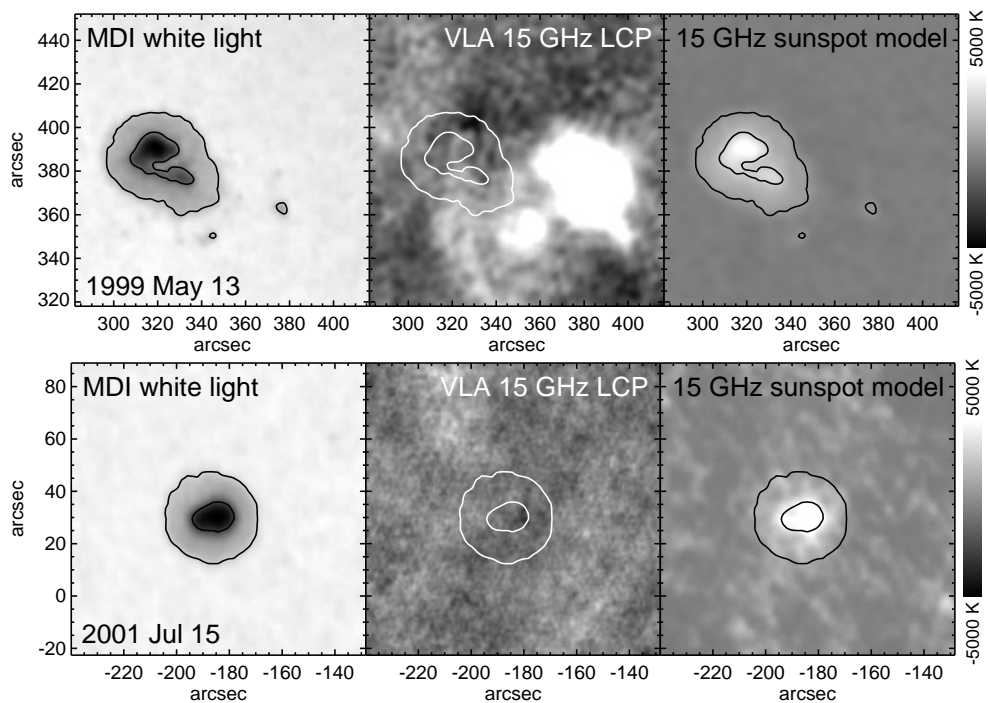}
  \caption{Two observations of sunspots made with the Very Large Array radio
telescope at 15 GHz. In both cases only the image in the sense of
circular polarization that minimizes any contribution from gyroresonance
emission (the ``ordinary'' mode) is shown. The upper panels compare a white light image from the
MDI instrument on the SOHO satellite (left), showing the umbra and
penumbra of a large sunspot observed on 1999 May 13, with a simultaneous
15 GHz image from the VLA in ``D'' configuration (middle, spatial resolution 5\arcsec).
The right panel shows what the VLA would have seen if the only emission
were from the sunspot, i.e., the white--light image scaled linearly
so that the darkest part of the umbra is
6000 K hotter than the surrounding atmosphere at 15 GHz.
The lower panels show another comparison for a sunspot observed on 2001
July 15 when the VLA was in ``C'' configuration (resolution 2\arcsec).
Because the VLA ``C'' configuration is not sensitive to large
structures, the model image in this case does not recover all the enhanced
emission from the penumbra, but the umbra is well recovered.
The actual and model radio images are displayed with a brightness
temperature range of
-5000 to 5000 K relative to the background quiet--Sun level (the VLA is
an interferometer and does not measure the absolute level). Contours
mark sunspot umbrae and penumbrae.}
    \label{fig9}
\end{figure*}

The magnetic field strength in the umbra can also affect
the predictions
of the models at millimeter wavelengths through its influence on the bremsstrahlung opacity.
Therefore we also studied the dependence of sunspot brightness
on the photospheric magnetic
field. In Fig.~\ref{fig7} we show an example of normalized brightness
temperature as a function of magnetic field strength at the photospheric level
for wavelengths of 1, 3, 6 and 10~mm derived using the model of
\citet{Fontenla2009}. For mm
wavelengths up to 6~mm (crosses, asterisks and diamonds in Fig.~\ref{fig7})
the magnetic field influences the model brightness by less than 0.5\% even
for field strengths of 3500~G. Crucially, a stronger field increases the
brightness of the umbra, so that we expect the discrepancy with the 3~mm
data to be even larger if the 2000~G field strength obtained by MDI is too low.
Starting from 10~mm the difference in
brightness between sunspots with different magnetic fields becomes more
significant. Thus, for a strong sunspot with photospheric magnetic field of
3500~G the model of \citet{Fontenla2009} predicts that the brightness
at 10~mm should be 3\% higher than for the sunspot analyzed
in this work (2000~G).
Longer wavelengths demonstrate a stronger dependence of sunspot
brightness on magnetic field, but this can be difficult to measure due
to the presence of a significant contributions from other sources,
discussed in the next section.

\section{Umbrae at centimeter wavelengths}\label{section4}

One consequence of the umbral brightness analysis at sub-mm and mm
wavelengths is that we
expect the radio umbra to change its appearance from darker than the quiet
Sun at short wavelengths to brighter than the quiet Sun at longer wavelengths.
The turnover wavelength beyond which (i.e., at longer wavelengths)
an umbra should appear brighter
than the quiet Sun depends on the exact temperature distribution in the
model and can be used as an independent test for existing sunspot
models. For the models considered here it ranges from 2.5~mm in the case of
\citet{Fontenla2009} to 0.3~mm in the \citet{Maltby1986} model (see Fig.~\ref{fig5}). As
reported in Sect.\ref{section3}, none of the models investigated has the turnover
wavelength in agreement with the 3.5~mm measurements, which show the umbra
to be cooler than the quiet Sun at this wavelength, whereas
all the models predict a brighter umbra at wavelengths longer than 3~mm.

The difference between the predictions of the various models tends to increase rapidly towards longer wavelengths, making it attractive to observe at, say, 10~mm. However, it is more difficult to
measure the umbral brightness--temperature offset at centimeter
wavelengths because of the presence of additional contributions to the
brightness temperature that are negligible at millimeter wavelengths: thus
dense material in the solar corona above the chromosphere contributes
optically thin thermal emission (brightness temperature $\propto\
\lambda^2$) that can reach tens of thousands of K, and gyroresonance
emission from strong magnetic fields in the umbra can reach coronal
brightness temperatures ($>$ 10$^6$ K) at microwave wavelengths if the
field strength is sufficiently high.
Gyroresonance opacity is prominent only at low harmonics (2 or 3) of the
electron gyrofrequency $\Omega_B\,=\,2.8\times 10^6 B$ Hz,
with $B$ measured in G. The fact that the observed
field strength rarely exceeds 4000~G (see, however, \citet{livingston}, \citet{Noort}, for exceptions) means that coronal gyroresonance emission is seldom seen above 20~GHz.

There have been a number of observations of large sunspots carried out with the
Very Large Array (VLA) at 15~GHz (2~cm wavelength) that are suitable to
investigate the umbral brightness temperature at centimeter
wavelengths.  At this frequency bright gyroresonance emission is
often present, if at all, only in the sense of circular
polarization that corresponds to the gyration motion of electrons about
the electric field (the magnetoionic ``extraordinary'' mode), because
the other polarization (the ``ordinary'' mode) requires a lower
harmonic to produce significant opacity and therefore requires a higher magnetic
field strength (2700 G for the 2nd harmonic) that is usually not present in the corona \citep[e.g., see][]{white97}.

Figure \ref{fig9} shows two examples of VLA observations of sunspots at
15 GHz in the ordinary--mode polarization (in both
cases bright gyroresonance sources are seen over the umbra in the opposite
polarization). In the sunspot observed on May 13, 1999,
optically thin thermal emission from dense coronal
gas can be seen to the west-south-west of the sunspot with a brightness
temperature of order 25000 K (saturated in the figure). Fainter coronal
contributions (1000-2000 K) can be seen to the south-west and east of
the spot. These regions of emission match the distribution of denser
coronal material
seen in Extreme--Ultraviolet Imaging Telescope (EIT) Fe XII 195 \AA\
images from the SOHO satellite.
There is no clear bright or dark feature corresponding to the
umbra of the sunspot. The noise level in regions away from the sunspot and
from the bright coronal emission is 900 K.
The spot of 2001 July 15 has no bright coronal emission near it (confirmed
by inspection of the EIT 195 \AA\
image): there is a small dark feature coincident with,
but much smaller than the umbra that is 3000 K below the background
level. The noise level in the VLA image is 800 K. Since the latter
observation was carried out in the VLA's ``C''-configuration which is
not sensitive to large sources, to confirm that the lack of an umbral
feature is not an artefact of the observing conditions we modelled what
the VLA would have seen in each case by assuming
that the umbra should be 6000 K (a lower limit, from Fig.~\ref{fig10})
brighter than the surrounding quiet Sun at 2 cm wavelength. We linearly
scaled the corresponding white-light image accordingly, used it to
generate model visibilities matching those used in the VLA observations,
and then mapped those data exactly as for the real data. The results are
shown in the right--hand column of Fig.~\ref{fig9}, and they confirm
that the absence of umbral features is not due to the observing
conditions. The ``C''-configuration model does indeed suggest that some
of the penumbral emission is resolved out, but the bright umbral feature
is recovered.

These examples are typical of VLA observations of sunspots at 15 GHz:
when other contributions are sufficiently small so as not to obscure the
umbra, the radio images at this frequency generally do not show a
distinct feature (bright or dark) corresponding to the umbra, with an
upper limit on the difference of order
1--2 thousand K. This result is again inconsistent with the umbral
atmosphere models that predict that umbrae should be much brighter than the
quiet Sun at centimeter wavelengths.

Once more among the considered models it is the model of \citet{Severino} that is in closest agreement with these data.
However, we note that the inference here
rests on the assumption that the brightness temperature surrounding the
umbra in these radio images is truly representative of the quiet Sun and
therefore corresponds to our reference model: given the additional
contributions to brightness temperature at centimeter wavelengths, and
the fact the the VLA field of view at 15 GHz is only 3\arcmin, it is
possible that the regions adjoining the umbra in our images are not
truly representative of the quiet Sun. This should not be an issue in
the case of the 2001 July 15 spot, which was
isolated with very little coronal emission around it.
However, pending better data, we emphasize that
the results of this section are not yet definitive.


\section{Conclusions}

Millimeter brightness observations impose strong constraints on temperature
and density stratifications of the sunspot atmosphere, in particular on the
location and depth of the temperature minimum and the location of the
transition region. Current submm/mm observational data suggest that
existing sunspot models, including the models of \citet{Avrett1981}, \citet{Maltby1986}, \citet{Obridko},
\citet{socas} and \citet{Fontenla2009}, fail to reproduce the millimeter
observations and therefore are incomplete descriptions of the
umbral atmosphere from the temperature minimum through the chromosphere up to
the transition region. Only the model of \citet{Severino} lies within the error bars at all the wavelengths considered here.

From the analysis of the existing models we can conclude that a successful
model that is in agreement with submillimeter and millimeter umbral
brightness should have an extended or/and deep temperature minimum (3000~K or
below) as, for instance, in the models of \citet{Severino} and \citet{Fontenla2009}. 

Most atmospheric models are based on one-dimensional static
atmospheres that try to reproduce observed umbral spectra. However, according to J.~Fontenla (priv. comm.)
there is an obvious reason why such 1D models will have difficulty. The atmosphere of a sunspot umbra is
strongly irradiated from the sides by the penumbra in the walls
produced by the Wilson depression, and this will affect the observed spectra.
An example is the fact that Lyman-$\alpha$ has centrally--peaked
profiles in the umbra but centrally--reversed profiles elsewhere
\citep[e.g.,][]{Tian09}: centrally--peaked profiles are expected if the
umbral Ly$\alpha$ emission is dominated by scattered and redistributed emission
coming in from the penumbral walls, rather than emission intrinsic to
the umbral atmosphere.
One-dimensional models cannot handle the complication introduced by
scattering from a source with a different atmospheric structure.
This implies that umbra is probably
cooler and less dense than the current 1D models indicate, which is consistent with the fact that the radio data show
the depression continuing to longer wavelengths than the 1D models suggest.

Due to the spatial resolution limit of 12\arcsec\ in the BIMA
observations used here we are not able to fully
resolve the umbra and cleanly separate it from penumbra. Therefore the results
obtained in this work are preliminary. A detailed study of the appearance
of sunspot umbrae at mm waves requires significantly higher spatial
resolution. Furthermore, good wavelength coverage is needed for accurate
diagnostics of the turnover wavelength, which is, in turn, required for the successful modeling of the
sunspot atmospheric temperature structure based on mm-wavelength data. We place high expectations on
the Atacama Large Millimeter/Submillimeter Array (ALMA), which has
commenced Early Science observations. The current instrument wavelength range covers
0.4 to 3.6~mm and the field of view ranges from 8.5\arcsec\ at the
shortest wavelengths to 72\arcsec\ at the longest wavelength.
At the shortest wavelengths the single pointing field of view will be too small for umbral observations and the mosaicing observing mode will be essential to
cover a sunspot together with surrounding quiet-Sun areas. With sub-arcsecond resolution and up to 66 antennas, ALMA will be an extraordinarily powerful instrument for studying the three-dimensional thermal structure of
sunspots at chromospheric heights.

\begin{acknowledgements} We thank Gene Avrett and Juan Fontenla for valuable discussions and Giuseppe Severino for kindly providing his model.
\end{acknowledgements}

\end{document}